\documentclass{article}
\pdfoutput=1
\usepackage{ComplexSystems}
\usepackage{graphics}
\usepackage{color}
\usepackage{hyperref}
\usepackage[english]{babel}
\usepackage[latin1]{inputenc}
\usepackage[T1]{fontenc}
\usepackage{amssymb}
\usepackage{amsmath}
\usepackage{booktabs}
\usepackage{tikz}
\usetikzlibrary{external}
\usetikzlibrary{positioning, decorations.pathmorphing, calc, scopes}
\tikzset{
  >=latex}                      
\definecolor{gray1}{gray}{.3}
\definecolor{gray2}{gray}{.65}
\definecolor{gray3}{gray}{.85}

\def\br{\linebreak[0]}          

\def\etal{\textit{et al.}}

\def\qqtext#1{\qquad\text{#1}\qquad}
\def\qtext#1{\quad\text{#1}\quad}

\def\ci#1{\includegraphics{ci/#1}} 

\def\Z{\mathbb Z}

\def\pr{\mathrm{pr}}                   
\def\cl{\mathop{\mathrm{cl}}\nolimits} 
\def\ovl#1{\langle #1 \rangle}         

\def\set#1{\{\, #1 \,\}}

\def\abs#1{\mathopen|#1\mathclose|}

\let\epsilon=\varepsilon
\let\phi=\varphi
\def\phiglobal{\phi_\mathrm{global}}

\def\op{{\oplus}}
\def\om{{\ominus}}

\def\rea{\rightarrow}

\def\wl{\overleftarrow{w}}
\def\wr{\overrightarrow{w}}

\begin{document}
\title{A Language for Particle Interactions in Rule 54 and Other
  Cellular Automata}
\author{\authname{Markus Redeker}\\[2ex]
  \authadd{Hamburg, Germany}\\
  \authadd{markus2.redeker@mail.de}}

\markboth{Complex Systems}{A Language for Particle Interactions}
\maketitle

\begin{abstract}\noindent
  This is a study of localised structures in one-dimensional cellular
  automata, with the elementary cellular automaton Rule 54 as a
  guiding example.

  A formalism for particles on a periodic background is derived,
  applicable to all one-dimensional cellular automata. One can compute
  which particles collide and in how many ways. One can also compute
  the fate of a particle after an unlimited number of collisions --
  whether they only produce other particles, or the result is a
  growing structure that destroys the background pattern.

  For Rule 54, formulas for the four most common particles are given
  and all two-particle collisions are found. We show that no other
  particles arise, which particles are stable and which can be
  created, provided that only two particles interact at a time. More
  complex behaviour of Rule 54 requires therefore multi-particle
  collisions.
\end{abstract}

\section{Introduction}

This article is part of a project to develop a higher-level language
for the dynamical behaviour of cellular automata. In the current
investigation we search for an intermediate-level description of the
elementary cellular automaton Rule 54, in order to learn how to handle
periodic background structures and simple particle interactions. The
investigation leads to further streamlining and an extension of the
existing formalism.\footnote{This article started as an extension of
  \cite{Redeker2010a}, but has now grown considerably and is
  completely rewritten.}

The formalism is called \emph{Flexible Time}. It was introduced in
\cite{Redeker2010} and further developed in \cite{Redeker2013a}.
Flexible Time makes it possible to ``calculate'' with the localised
structures in a cellular automaton and to determine their development
over time. The structures in Flexible Time resemble the way in which a
human observer views an evolution diagram of a cellular automaton
(like Figure~\ref{fig:random}):
\begin{figure}[ht]
  \centering
  \ci{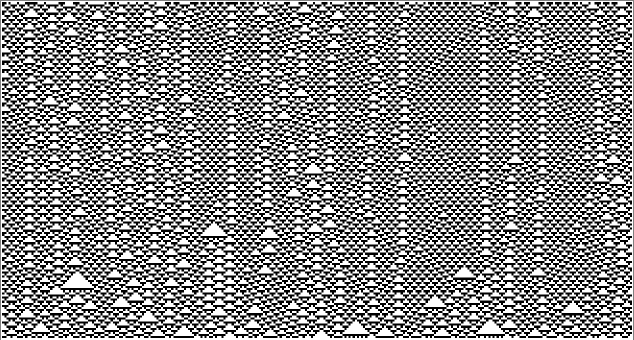}
  \caption{Development of a random initial configuration under Rule
    54. Time runs from bottom to top.}
  \label{fig:random}
\end{figure}
by grouping the states of cells from different times and places to a
single pattern in space-time.

Rule 54 is an elementary cellular automaton that was first
investigated in detail by Boccara \etal\ \cite{Boccara1991}. When
evolving from random initial configurations, it develops a simple
background pattern with a small number of interacting particles that
move on this background. While it has not been shown to be
computationally universal, it can at least evaluate Boolean
expressions \cite{Ju'arezMart'inez2006a}. So it is a rather simple
system (but not too simple) and therefore an ideal test object for a
formalism that is still under development.

The right methods to handle large complex structures must still be
found. I ask here new questions about the behaviour of Rule 54, and
Flexible Time must ``learn'' how to handle them. As a result, there
are differences and extensions of the formalism in this article that
were not present in \cite{Redeker2013a}. I will point them out and
review them at the end.

\paragraph{Context} Researchers on cellular automata have developed a
number of concepts to describe the localised structures that arise in
a cellular automaton.

The oldest named structures must be the particles (also called gliders
or signals) and their collisions. This goes back at least to Zuse
\cite{Zuse1967}, whose cellular automaton simulates idealised physical
particles. Particle-based research has continued since then, with
Cook's construction of a universal computer in Rule 110 as its most
spectacular result~\cite{Cook2004}.

This rule, and Rule 54, belonged also to those rules in which a stable
periodic background pattern occured; it was called ``ether'' by Cook.

For Rule 54, the starting point was the work by Boccara \etal\
\cite{Boccara1991}; they identified the most common particles that
arise from random initial configurations, described their interactions
and gave them the names that are still used. This research was later
continued by the group around McIntosh
\cite{Ju'arezMart'inez2006a,Martinez2014,Ju'arezMart'inez2008}, who
found more complex particles and interactions.

The descriptions of these particles were mostly given by pictures and
by a simple symbolism that showed which particle collides with which.
But, especially to find general theorems about cellular automata, more
abstract representations were developed too.

There is a more detailed investigation of particles and what they can
achieve \cite{Durand-Lose2008,Mazoyer1999a}. For Rule 110 there is an
approach for the systematic specification of initial configurations
with interacting gliders \cite{Ju'arezMart'inez2008}, and to express
the behaviour of the cellular automaton through a block substitution
system \cite{Seck-Tuoh-Mora2010}.

There are also the approaches by Hordijk \etal\ \cite{Hordijk2001} and
by Martin \cite{Martin2000}, who use properties of the background and
draw conclusions about the particles and particle interactions that
are possible. More general, the cellular automaton is subdivided in
``regular'' regions and the boundaries between them
\cite{Eloranta1993,Hanson1995,Jen1990,Li1992,Pivato2007}; the
boundaries move, often in an almost random fashion, and are thus a
generalisation of the more straight-moving particles.

Other approaches view the evolution of the cellular automaton as
two-dimensional, with one space and one time dimension. The cellular
space-time is then subdivided into finite patches that represent
e.\,g.\ a piece of the background or a collision between particles.
The theory of cellular automata then becomes a special tiling problem.
One can do this in a more informal way, like McIntosh and Martínez
\cite{Ju'arezMart'inez2001}, or develop a complex formal theory around
it, as Ollinger and Richard \cite{Ollinger2007,Ollinger2009} do it.
(This approach is closest to the work described here.)

\paragraph{Overview} After an introductory section about cellular
automata and Rule 54, Section~\ref{sec:local-interactions}
recapitulates the work in \cite{Redeker2013a}, as far as it is
relevant for the present work. At its end, a representation of Rule 54
as a ``reaction system'' (defined below) is shown, the same that was
derived in \cite{Redeker2013a}. In Section~\ref{sec:understanding} we
then find a way to compress this and similar systems, and we use the
compressed reaction system to understand the local behaviour of Rule
54 better. Section~\ref{sec:triangles-ether} then turns to larger
patterns and describes the triangular structures in Rule 54 and the
stable background pattern that is formed by them. Then, in
Section~\ref{sec:particles}, the four kinds of particles found by
Boccara \etal\ \cite{Boccara1991} are represented in Flexible Time,
together with the collision between the particles. A summary follows
in Section~\ref{sec:conclusion}.

\section{Cellular automata and Rule 54}
\label{sec:cellular-automata}

\subsection{Elementary cellular automata}

Rule 54 is an one-dimensional cellular automaton, more specifically an
\emph{elementary cellular automaton}. This kind of cellular automata
was made popular by Stephen Wolfram \cite{Wolfram1984}, who also
introduced the system of code numbers from which Rule 54 got its name.

One-dimensional cellular automata are dynamical systems with discrete
time. The state of such an automaton is called a \emph{configuration}.
It consists of an infinite sequence of simpler objects, the
\emph{cells}. The state of each cell is an element of a finite set
$\Sigma$; the configuration at time $t$ is therefore a function $c_t
\colon \Z \to \Sigma$. We write $\Sigma^\Z$ for the set of
configurations; $c_t(x)$ is then the state of the cell at position $x$
at time~$t$.

The \emph{evolution} of the automaton is then a sequence $(c_0, c_1,
c_2, \dots)$ of configurations that follow a common rule that is
described below in~\eqref{eq:global}. While the sequence here starts
at time 0, we will also accept other starting times.)

An elementary cellular automaton is a one-dimensional cellular
automaton with two states and a three-cell neighbourhood. The set of
states is $\Sigma = \{0, 1\}$, and its behaviour is given by its
\emph{local transition rule}
\begin{equation}
  \phi\colon \Sigma^3 \rightarrow \Sigma\,.
\end{equation}
This is the function with which the configuration $c_{t+1}$ is
computed from its predecessor $c_t$. To do this, we apply $\phi$ to
every three-cell neighbourhood of $c_t$, and the result is the next
state of the middle cell:
\begin{equation}
  \label{eq:global}
  c_{t+1}(x) = \phi(c_t(x - 1), c_t(x), c_t(x + 1))
  \qquad\mbox{for all $t$, $x \in \Z$}.
\end{equation}
The function $\phi$ defines then a \emph{global transition rule}
$\phiglobal$: It is the function that maps the configuration $c_t$ to
its successor $c_{t+1}$ according to \eqref{eq:global}.

The transition rule~\eqref{eq:global} is also called a rule of
\emph{radius} 1, because only the $c_t(y)$ with $\abs{x - y} \leq 1$
contribute to $c_{t+1}(x)$. Rules with other radii are defined
similarly.

\subsection{Rule 54}

Rule 54 has a left-right symmetric transition rule,
\begin{equation}
  \label{eq:def54}
  \phi(s) =
  \left\{
    \begin{array}{r@{\quad}l}
      1&\mbox{for $s \in \{(0, 0, 1), (1, 0, 0), (0, 1, 0), (1, 0, 1)\}$,}\\
      0&\mbox{otherwise.}
    \end{array}
  \right.
\end{equation}
The rule is easier to remember in form of the following slogan
\cite{Redeker2013a},
\begin{quote}
  \label{par:slogan}
  ``$\phi(s) = 1$ if $s$ contains at least one 1, except if the cells
  in state 1 touch.''
\end{quote}
Here we say that two cells ``touch'' if they are direct neighbours.
Thus the two cells in state 1 touch in the neighbourhood $(1, 1, 0)$,
but not in the neighbourhood $(1, 0, 1)$.

\begin{figure}[ht]
  \centering
  \ci{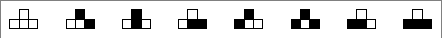}
  \caption{Rule icon for Rule 54.}
  \label{fig:rule-icon}
\end{figure}
Figure~\ref{fig:rule-icon} shows how the neighbourhoods influence the
next state of the central cell. White squares are in state 0, black
squares are in state 1, and the time runs upwards. This is also our
convention in the other diagrams, even if white and black may also
become dark and bright grey in the parts of the diagram that are
less emphasised.

\section{Flexible Time}
\label{sec:local-interactions}

\subsection{Situations}
\label{sec:situations}

We need a means to describe and understand the interactions of gliders
and other patterns under Rule 54. Flexible Time was developed in
\cite{Redeker2013a} for this task. The motivation was that it is
easier to find patterns in the evolution of cellular automata if one
works with structures that involve the states of cells at different
times. These structures are called here \emph{situations}.

They generalise the finite sequences of cells that are part of the
configurations $c_t$ described above. In order to express e.\,g.\ that
$c_t(0) = c_t(1) = 0$ and $c_t(2) = 1$, one would often write that the
subsequence of $c_t$ that begins at cell position 0 is 001. Situations
generalise this notation. They may extend not only over space but also
over time. To write them, we use additional symbols that express a
jump in spacetime.

Under Rule 54, situations are written as sequences of the symbols 0,
1, $\ominus_i$ and $\oplus_i$, for $i \in \{1, 2\}$. The intended
interpretation can most easily be described in terms of instructions
to write symbols on squares in a grid. The squares are labelled by
pairs $(t,x) \in \mathbb Z^2$; $x$ is the position of a cell and $t$ a
time step in its evolution. The writing rules are:
\begin{itemize}
\item Start reading at the first symbol. For writing, place the cursor
  at square $(0,0)$ of the grid.
\item If the cursor is at $(t, x)$ and the current symbol is
  \begin{itemize}
  \item an element of $\Sigma$, write it down and move the cursor one
    square to the right, to $(t, x + 1)$,
  \item $\ominus_i$, move the cursor to $(t - 1, x - i)$,
  \item $\oplus_i$, move the cursor to $(t + 1, x - i)$.
  \end{itemize}
  Then continue with the next symbol.
\item \emph{No overwriting}: One cannot write different symbols on the
  same square.
\end{itemize}

To get an example for such a writing process, let us set for a moment
$\Sigma = \{ 0, 1, 2, 3 \}$ and look at the situation $01 \op_1 23$.
First, the cell states 0 and 1 are written to the squares $(0, 0)$ and
$(0, 1)$. The cursor is then at square $(0, 2)$. Now the symbol
$\op_1$ moves the cursor to $(1, 1)$. The following symbols 2 and 3
are then written to the squares $(1, 1)$ and $(1, 2)$, leaving the
cursor at $(1, 3)$. The result is then the following grid:
\begin{equation*}
  \tabcolsep=0.7ex
  \color{gray2}
  \def\b#1{\textcolor{black}{#1}}
  \def\r{\b{\rule[.5ex]{1.5em}{.7pt}}}
  \begin{tabular}{r||c|c|c|c|c|c|c}
    $t=1$  &    &         &     & \b2 & \b3 & \rlap\r & \\ \hline
    $t=0$  &    & \llap\r & \b0 & \b1 &     &         & \\ \hline \hline
    $x={}$ & -2 & -1      & 0   & 1   & 2   & 3       & 4
  \end{tabular}
\end{equation*}
The horizontal rules mark the beginning and end of the symbol
sequence, or, more exactly, the squares left of the starting point and
right of the end point of the state sequence. Similar lines will later
appear in the illustrations.

Now we need to express this construction in a mathematical form. We
will use two-dimensional coordinates and call a coordinate pair $(t,
x) \in \Z^2$ a \emph{space-time point}. A pair $(p, \sigma) \in \Z^2
\times \Sigma$ is a \emph{cellular event}. The event $((t, x),
\sigma)$ provides the information ``At time $t$, the cell at position
$x$ is in state $\sigma$''. We will usually write them $[t, x]\sigma$
or $[p]\sigma$ for better readability. A situation is then a sequence
of cellular events, together with the final cursor position: $s =
(([p_0] \sigma_0, \dots, [p_{n-1}] \sigma_{n-1}), p_n)$. For the final
cursor position of $s$ we write $\delta(s)$, the \emph{size} of $s$.
This means that we have in our example $\delta(s) = p_n$.

In a situation, the sequence of the cellular events is significant,
and the size too, since they make algebraic operations possible. In
many cases, we want to ignore however this information: Then we will
use the \emph{cellular process} that belongs to a situation; it is
simply the set of its cellular events. The cellular process of a
situation $s$ is written $\pr(s)$. In our example, with $s = 01 \op_1
23$, we have therefore
\begin{equation*}
  s = (([0, 0] 0, [0, 1] 1,\br [1, 1] 2, [1, 2] 3), (1, 3))\,.
\end{equation*}
This means that $\delta(s) = (1, 3)$ and $\pr(s) = \{ [0, 0] 0,\br [0,
1] 1, [1, 1] 2, [1, 2] 3 \}$.

Usually we will not need this explicit form, since situations are
meant to make this unnecessary. It helps us however to explain the
``no overwriting'' rule above. This rule concerns expressions like $01
\op_1 2 \om_1 3$, where the cursor reaches the same point twice. If it
were a situation, its cellular process would be $\{ [0, 0] 0,\br [0,
1] 1,\br [1, 1] 2, [0, 1] 3 \}$. This would provide contradicting
information about the space-time point $(0, 1)$: At time 0, is the
cell at position 1 in state 0 or 3? The overwriting rule prevents this
problem.

The most important algebraic property of situations is that they can
be multiplied. The product of $s_1$ and $s_2$ is found by first
writing $s_1$ and then, with the cursor at $\delta(s_1)$, writing
$s_2$. The resulting product is written $s_1 s_2$, but due to the
overwrite rule, it may not always exist.

More complex terms of situations are defined in the usual way: $s^2$
is the result of writing $s$ twice, and so on. The \emph{Kleene
  closure} of a situation $s$ is the set
\begin{equation}
  s^* = \set{s^k \colon k \geq 0}\,.
\end{equation}
The Kleene closure always contains the \emph{empty situation}, which
is written~$[0]$.

In Flexible Time, situations are used to express the evolution of a
cellular automaton. But in order to understand how this is done, we
first have to look at the way in which the evolution of a cellular
automaton is expressed by cellular processes.

\subsection{Evolution Expressed with Cellular Processes}

In a similar way to that in which a configuration $c_0 \in \Sigma^\Z$
can be the starting point of an evolution $(c_0, c_1, c_2, \dots)$, a
cellular process $\pi$ can be extended to a larger process $\cl \pi$,
its \emph{closure}.
\begin{figure}[ht]
  \centering
  \ci{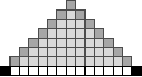}
  \caption{A process and its closure.}
  \label{fig:example-closure}
\end{figure}
Figure~\ref{fig:example-closure} shows how this is meant for the
initial configuration $\pi = \pr(10^{13}1)$. The cellular events of
the original process $\pi$ are displayed in black and white; together
with the the events in grey they form the process $\cl \pi$. Each
horizontal row in the diagram contains the events that belong to a
specific time step. We see that the diagram becomes smaller at the
top; this means that as time progresses, fewer cell states can be
deduced from the information given by the initial process $\pi$.

To motivate the exact definition of the closure, we first express the
configurations of the cellular automaton and their evolution in terms
of cellular processes. This will then allow us to generalise the
definition of evolution to processes that do not correspond to
configurations.

Let now $c$ be the configuration of a cellular automaton. We define
the \emph{embedding} of $c$ at time $t$ to be the process
\begin{equation}
  \label{eq:embedding}
  \eta_t(c) = \set{ [t, x] c(x) \colon x \in \Z}.
\end{equation}
A kind of inverse of the function $\eta_t$ is the concept of
\emph{time slices}. The time slice at time $t$ of a process $\pi$ is
the process
\begin{equation}
  \label{eq:time-slice}
  \pi^{(t)} = \set{ [t, x] \sigma \colon x \in \Z }.
\end{equation}
The time slice is a process and not a configuration because
$\pi^{(t)}$ must exist for all processes, not just for embeddings of
configurations.

With these concepts, the cellular process that belongs to the
evolution sequence $(c_o, c_1, c_2 \dots)$ is $\gamma = \bigcup_{t
  \geq 0} \eta_t(c_t)$. It has the time slices $\gamma^{(t)} =
\eta_t(c_t)$, which represent the configurations $c_t$. The process
$\gamma$ must then be the closure of $\eta_0(c_0)$.

A time slice $\pi^{(t)}$ of an arbitrary process is then understood as
partial knowledge about the state of a cellular automaton at time $t$.
In order to determine the state of the automaton at time $t + 1$, we
take all configurations that are compatible with this knowledge,
evolve them for one time step and accept only the states of those
cells about which all configurations agree. The result is the cellular
process
\begin{equation}
  \label{eq:determined}
  \Delta_t(\pi) = \bigcap \set{
    \eta_{t+1}(\phiglobal(c)) \colon \eta_t(c) \supseteq \pi^{(t)} }
\end{equation}
of those events that are \emph{determined} by $\pi{(t)}$. The cellular
events of which it consists all belong to time $t + 1$.

We can now easily check that the process $\gamma$ has the property
that $\gamma^{(t)} = \Delta_t(\gamma)$ for all $t > 0$. Every time
slice, except the first, can be computed from the previous one. Only
$\gamma^{(0)}$, which represents the initial configuration, must still
be handled separately.

This inconvenience in resolved in the full definition of the closure.
In it, the initial process no longer needs to be the embedding of a
configuration. This is possible because it is now split into time
slices and then added piece-wise to the partial results of the
computation.

\begin{definition}[Closure {\cite[Def.~3.10]{Redeker2013a}}]
  \label{def:closure}
  Let $\pi$ be a cellular process for which there is a time $t_0 \in
  \Z$ such that $\pi^{(t)} = \emptyset$ for all $t < t_0$.

  If there is a process $\gamma$ with the property that
  \begin{equation}
    \label{eq:closure}
    \gamma^{(t)} =
    \begin{cases}
      \Delta_t(\gamma) \cup \pi^{(t)} & \text{for $t \geq t_0$,} \\
      \emptyset & \text{for $t < t_0$,}
    \end{cases}
  \end{equation}
  then we write $\gamma = \cl \pi$ and say that it is the
  \emph{closure} of $\pi$.
\end{definition}
It is easy to see that the choice of $t_0$ has no influence on $\cl
\pi$.

We can now see that the set $\gamma$ that was defined above
satisfies~\eqref{eq:closure} if we set $t_0 = 0$ and $\pi =
\eta_0(c_0)$: Then we have $\gamma^{(t)} = \emptyset$ for $t < 0$,
$\gamma^{(0)} = \eta_0(c_0)$, and $\gamma^{(t)} = \Delta_t(\gamma)$
for $t > 0$, and indeed $\gamma = \cl \eta_0(c_0)$.
Definition~\ref{def:closure} is thus a generalisation of the
transition rule~\eqref{eq:global} to cellular processes.

Not to all cellular processes, however. One of the requirements of
Definition~\ref{def:closure} is that $\gamma$ must be a process, and
this can easily be broken. All we need is conflicting information in
$\Delta_t(\gamma)$ and $\pi^{(t)}$: If there is a time step $t$ at
which there is an event $[t, x] \sigma \in \Delta_t(\gamma)$ and
another event $[t, x]\tau \in \gamma^{(t)}$ with $\sigma \neq \tau$,
then $\gamma^{(t)}$ is no cellular process, and neither is~$\gamma$.

We will however introduce in the next section a class of situations
whose cellular processes all have a closure: They will then be used to
describe the evolution of cellular automata in an economical way.

\subsection{Reactions}

The evolution of a cellular automaton is represented in Flexible Time
by \emph{reactions}. We will say that there is a reaction between two
situations $a$ and $b$ if the situation $b$ consists only of events
that are determined by of the events of $a$. They belong to the future
of $a$, so to speak.

\begin{figure}[ht]
  \centering
  \def\arraystretch{1.5}
  \begin{tabular}{c}
    \ci{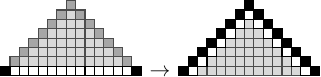} \\
    $\llap{$1 0^{13} 1$} \rea \rlap{$(10\op)^7 1 (\om 01)^7$}$
  \end{tabular}
  \caption{A reaction under Rule 54.}
  \label{fig:example-reaction}
\end{figure}
Figure~\ref{fig:example-reaction} shows a reaction. On the left side
we see the process of the situation $a = 1 0^{13} 1$, together with
its closure. As in Figure~\ref{fig:example-closure}, the events of
$\pr(a)$ are highlighted while the remaining cells of the closure are
displayed in grey. On the right side we see the same closure, but with
different events highlighted. This time they belong to the situation
$b = (10\op)^7 1 (\om 01)^7$. With these diagrams we therefore see
that the events of the process $b$ are determined by the process $a$.

The formal definition of reactions is then:
\begin{definition}[Reactions {\cite[Def.~4.8]{Redeker2013a}}]
  \label{def:reactions}
  Let $a$ and $b$ be two situations with
  \begin{equation}
    \label{eq:reaction}
    \cl \pr(a) \supseteq \pr(b)
    \qqtext{and}
    \delta(a) = \delta(b)\,.
  \end{equation}
  Then the pair $(a, b)$ is the \emph{reaction} from $a$ to $b$. It is
  usually written $a \rea b$.
\end{definition}
We will use the expression $a \rea b$ also as a proposition, meaning
that there is a reaction from $a$ to $b$. The symbol ``$\rea$'' then
specifies a relation, and as it is normal for relations, we can also
write longer chains of reactions, like $a \rea b \rea c$. One can
verify that if such a chain exists, then there is also a reaction $a
\rea c$.

Reactions are useful because they can be \emph{applied} to situations.
It can be shown \cite[Th.~4.11]{Redeker2013a} that if there are
situations $x$, $y$ and $a$ for which $\cl \pr(x a y)$ exists and if
there is a reaction $a \rea b$, then there is also a reaction $x a y
\rea x b y$. This reaction is then called the \emph{application} of $a
\rea b$ to $x a y$.

Now it is possible that there is also a reaction that can be applied
to $x b y$. We would then have a reaction $b' \rea c$ and two
processes $x'$ and $y'$ such that $x b y = x' b' y' \rea x' c y'$ and
therefore, by transitivity, also a reaction $x a y \rea x' c y'$. This
way application allows one to specify a large set of reactions by a
small set of ``generator reactions'', provided only that there is a
large enough set of situations to which they can be applied.

The result is a \emph{reaction system}. It is the foundation of all
calculations in Flexible Time.
\begin{definition}[Reaction System {\cite[Def.~4.13]{Redeker2013a}}]
  \label{def:reaction-system}
  Let $D$ be a set of situations and $R$ a set of reactions between
  them. We say that $R$ is a \emph{reaction system} with \emph{domain}
  $D$ if the following is true:
  \begin{enumerate}
  \item If $a \in D$, then $a \rea a$ is in $R$.
  \item If $a \rea b$ and $b \rea c$ are in $R$, then $a \rea c$ is in
    $R$.
  \item $R$ is closed under application of reactions to the situations
    in $D$.
  \end{enumerate}
\end{definition}

\begin{table}[t]
  \centering
  \caption{The local reaction system for Rule 54, long form.}
  \def\r{$&${}\rea}
  \tabcolsep=0pt
  \begin{tabular*}{\textwidth}{lrl@{\quad}rl}
    \toprule
    Gene\rlap{rating Slopes:} \\
    & $\om_1 00$, $1 \om_1 01$&, $1 \om_2 10$, $00 \om_2 11$,
    & $00 \op_1$, $01 \op_2 1$&, $10 \op_1 1$, $11 \op_2 00$. \\
    \midrule
    Reacti\rlap{ons:}
    & $   \om_1 000 \r   0\om_1 00$ & $000\op_1    \r 00\op_1 0$ \\
    & $   \om_1 001 \r   1\om_1 01$ & $100\op_1    \r 10\op_1 1$ \\
    & $  1\om_1 010 \r 111\om_2 10$ & $010\op_1 1  \r 01\op_2 111$ \\
    & $  1\om_1 011 \r 100\om_2 11$ & $110\op_1 1  \r 11\op_2 001$ \\
    & $  1\om_2 100 \r   1\om_1 00$ & $001\op_2 1  \r 00\op_1 1 $ \\
    & $  1\om_2 101 \r   1\om_1 01$ & $101\op_2 1  \r 10\op_1 1 $ \\
    & $ 00\om_2 110 \r 001\om_2 10$ & $011\op_2 00 \r 01\op_2 100$ \\
    & $ 00\om_2 111 \r 000\om_2 11$ & $111\op_2 00 \r 11\op_2 000$ \\[1.2ex]

    & $ 00 \r 00\op_1    \om_1 00$ & $   \om_1 00 \op_1    \r [0]$ \\
    & $ 01 \r 01\op_2 1  \om_1 01$ & $ 1 \om_1 01 \op_2 1  \r 1$ \\
    & $ 10 \r 10\op_1 1  \om_2 10$ & $ 1 \om_2 10 \op_1 1  \r 1$ \\
    & $ 11 \r 11\op_2 00 \om_2 11$ & $00 \om_2 11 \op_2 00 \r 00$ \\
    \bottomrule
  \end{tabular*}
  \label{tab:generator54}
\end{table}
We will now define a reaction system by specifying $D$ and a set $G
\subseteq R$ of generators; it is then extended by repeated
application and concatenation of reactions, as described above. The
system describes Rule 54; its derivation is described in detail in
Chapters 6 and 7 of \cite{Redeker2013a}.

The reaction system is summarised in Table~\ref{tab:generator54}. The
top of the table, entitled ``Generating Slopes'', specifies the domain
$D$ of $\Phi$. More specifically, it lists the neighbourhoods that a
$\ominus$ or $\oplus$ operator may have if it is is part of a
situation $s \in D$. The first entry, $\ominus_1 00$, specifies that a
$\ominus_1$ may occur in $s$ at the left of the term 00, the second
entry $1 \ominus_1 01$, that it may occur between a 1 (at its left)
and a 01 (at its right). No other possibilities exist since the
remaining entries refer to other operators. One can prove
\cite[Theorem 6.10]{Redeker2013a} that all situations in $D$ have a
closure. 

The bottom of Table~\ref{tab:generator54} contains the generating
reactions of $\Phi$. Its upper part (i.\,e.\ the middle of the whole
table) contains the reactions that involve a single $\ominus$ or
$\oplus$ operator. If we had only them, no reaction could have an
element of $\Sigma^*$ at its left side: Therefore we have at the
bottom left of the table a set of reactions that create a $\ominus$
and a $\oplus$ operator from an element of $\Sigma^*$. Their converses
are listed at the bottom right: reactions that destroy a $\ominus$ and
a $\oplus$ operator. All reactions of $\Phi$ are the results of
repeated applications of these four types of generators.

The arrangement of the reactions in Table~\ref{tab:generator54} has
also another purpose. It allows one to read off two important
subsystems of $\Phi$.

\begin{definition}[Slopes]
  \label{def:slopes}
  Let $R$ be a reaction system with domain $D$.

  The system $R_+$ (with domain $D_+$) of \emph{positive slopes}
  consists of the situations of $D$ that only contain $\oplus$
  operators and the reactions between these situations.

  The system $R_-$ (with domain $D_-$) of \emph{negative slopes}
  consists of the situations of $D$ that only contain $\ominus$
  operators and the reactions between these situations.
\end{definition}

In case of Rule 54, we can find the generators of $\Phi_-$ if we take
only the generating slopes at the right and the generator reactions at
the top right of the middle section in Table~\ref{tab:generator54}.
Similarly, $\Phi_+$ is represented by the slopes and reactions at the
top right of the table.

\paragraph{Details of the reaction system} We will now have a closer
look at the way in which the reaction system $\Phi$ represents Rule
54.

\begin{figure}[ht]
  \centering
  \ci{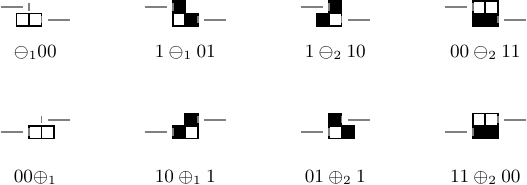}
  \caption{Generating slopes.}
  \label{fig:generators}
\end{figure}

We begin with the slopes. Figure~\ref{fig:generators} displays the
generating slopes for $\Phi$, first the negative slopes and then their
mirror images, the positive slopes.

In this and in later diagrams, the endpoints of the situations are
marked by horizontal lines. They represent the places where the
surrounding events would be expected if the slopes were parts of
larger situations. Or, in the interpretation of
Section~\ref{sec:situations}, the square at which the left horizontal
line ends is always one point left of the coordinate origin, while the
right horizontal line always begins at $\delta(s)$. The beginning of
the situation is also marked by the small vertical bar, which is
located at the left boundary of the square at the coordinate origin.

An important property of the generating slopes is that they trace the
boundaries of the closure.
\begin{figure}[ht]
  \centering
  \ci{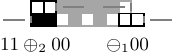}
  \caption{Generating slopes as boundaries of the closure.}
  \label{fig:boundary_slopes}
\end{figure}
We can see in Figure~\ref{fig:boundary_slopes} what this means. It
shows a situation, $110101000$, together with two generations of its
closure. We see at its left the slope $00 \oplus_2 11$ (the mirror
image of $11 \ominus_2 00$ in Figure~\ref{fig:generators}), and at its
right, the term $\ominus_1$, both in bolder colours. Note that the
situation $\ominus_1 00$ reaches over two time steps and has its
starting point directly at the right end of the second time step of
the closure. This is the way the slope terms trace the boundary of a
closure.

\begin{figure}[ht]
  \centering
  \ci{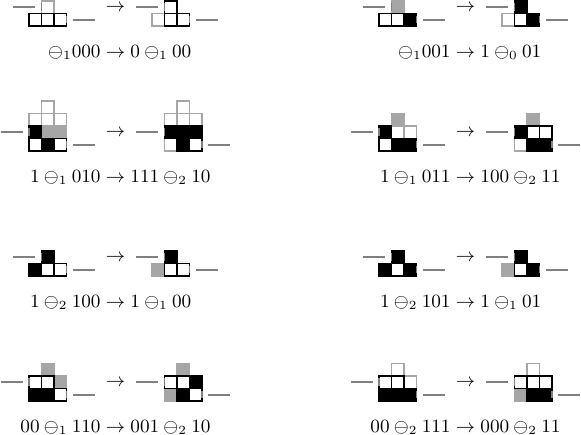}
  \caption{Reactions of $\Phi_-$ as motion towards the boundaries of
    the closure.}
  \label{fig:reactions-minus}
\end{figure}

\begin{figure}[ht]
  \centering
  \ci{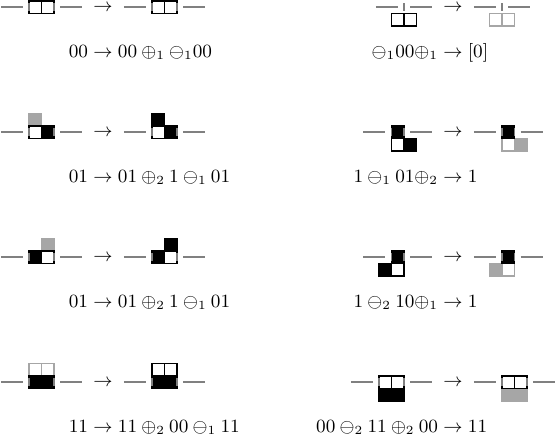}
  \caption{Reactions that generate and destroy slopes. The generator
    reactions are shown at the left, the destructors, right.}
  \label{fig:reactions-generate}
\end{figure}

The generator reactions of $\Phi_-$ are designed with the goal that
the reaction result consists of events near the right boundary of the
closure of the initial situation. (For $\Phi_+$ it is similar, with
left and right exchanged.) How this is done is shown in
Figures~\ref{fig:reactions-minus} and~\ref{fig:reactions-generate}.
They contain reactions of the form $a \rea b$ and display $\pr(a)$ and
$\pr(b)$ in relation to the closure of $\pr(a)$.
Figure~\ref{fig:reactions-minus} shows the generator reactions of
$\Phi_-$. In it, we see that the process of $b$ is always located more
to the right than $\pr(a)$ and that it touches the right boundary of
$\cl \pr(a)$. The reactions involve only two time steps, and one of
the $\ominus$ operators must always be present.
To get the system started from situations in $\Sigma^*$, we need the
reactions at the right side of Figure~\ref{fig:reactions-generate}.
Here we see reactions in which $\pr(b)$ completely fills the closure
of $\pr(a)$ and $b$ is a situation with both a $\oplus$ and a
$\ominus$ operator.

The converses of the reactions at the left side of
Figure~\ref{fig:reactions-generate} are shown at its right side:
Reactions $a \rea b$ in which $a$ contains one $\ominus$ and one
$\oplus$, while $b$ contains none. We can use them for cleanup, since
they remove pairs of neighbouring $\ominus$ and $\oplus$ operators.
The same manoeuvre is also possible in all other cases where a
$\ominus$ is left of a $\oplus$, and we get a result that for every
situation $a$ there is a reaction $a \rea b_+ b_-$ with $b_+ \in D_+$
and $b_- \in D_-$. If we start from $a$ and continue to apply the
generator reactions as long as possible, we can even enforce that
$b_+$ and $b_-$ trace the boundaries of $\cl \pr(a)$.

This was a summary of the content of \cite{Redeker2013a} as far as it
concerns Rule 54.

\section{Understanding the Reaction System}
\label{sec:understanding}

Up to now, the representation of Rule 54 in
Table~\ref{tab:generator54} looks complex and does not provide much
insight. This makes it difficult to do calculations about Rule 54
without always looking at the table. We will therefore develop a more
compact representation of the reaction system. The goal is to find
``slogans'' for it that are easy to remember, analogous to the slogan
for $\phi$ on page~\pageref{par:slogan}.

\subsection{A simpler Rule Table}

As a first simplification, we omit the indices from the $\oplus$ and
$\ominus$ operators. This is possible because the indices of the
operators are always determined by the environment. We can see from
the list of generating slopes in Table~\ref{tab:generator54} that if
$\ominus_i$ is followed by a $0$, then always $i = 1$, and if it is
followed by a $1$, then $i = 2$. A similar law is valid for
$\oplus_i$, and we can recover the indices of $\ominus$ and $\oplus$
from the equations
\begin{align}
  \label{eq:indices}
  \ominus 0 &= \ominus_1 0, &
  \ominus 1 &= \ominus_2 1, &
  0 \oplus &= 0 \oplus_1, &
  1 \oplus &= 1 \oplus_2\,.
\end{align}
This kind of abbreviation is possible in every reaction system,
because in a generating slope $u \ominus_i v$, the term $u \ominus_i$
is completely determined by~$v$.

For the same reason we can shorten the generator reactions by removing
common factors from their left and right sides. The generator
reactions of $\Phi_-$ all have the form $u \ominus v \sigma \rea u x
\ominus v'$, with a generating slope $u \ominus v$. When such a
reaction is applied to a situation $s$, there must be always a factor
$u$ to the left of $\ominus v$ in $s$. Therefore we can shorten these
generator reactions to the form $\ominus v \sigma \rea x \ominus v'$
and do not get new reactions when the shortened reactions are applied.

We then get four pairs of reactions as generators for $\Phi_-$:
\begin{subequations}
  \begin{align}
    \ominus 000 & \rea 0\ominus 00, & \ominus 010 & \rea 11\ominus 10, \notag\\
    \ominus 001 & \rea 0\ominus 01, & \ominus 011 & \rea 00\ominus 11, \\[1.5ex]
    \ominus 100 & \rea \ominus 00,  & \ominus 110 & \rea 1\ominus 00,  \notag\\
    \ominus 101 & \rea \ominus 01,  & \ominus 111 & \rea 0\ominus 01\,.
  \end{align}
\end{subequations}
They can be compressed further with the help of a new notation. For a
cell state $\sigma \in \Sigma$ we will write $\bar\sigma$ for the
\emph{complementary state}, such that $\bar 0 = 1$ and $\bar 1 = 0$.
Then we can write the following reactions, valid for all
$\sigma$,\footnote{The bottom left reaction has been shortened even
  more, it should have been $\ominus 10\sigma \rea \ominus 0\sigma$.}
\begin{subequations}
  \begin{align}
    \ominus 00\sigma & \rea \sigma \ominus 0 \sigma, &
    \ominus 01\sigma & \rea \bar\sigma \bar\sigma \ominus 1 \sigma, \\
    \ominus 10       & \rea \ominus 0,               &
    \ominus 11\sigma & \rea \bar\sigma \ominus 1 \sigma\,.
  \end{align}
\end{subequations}
Written in this form we will analyse the reaction system and show what
the generator reactions actually mean. But before we can do this, we
must see how to simplify the rest of Table~\ref{tab:generator54}.

The reactions at the bottom of the table can be brought easily to a
common form, when we define the set $G_- = \{ \ominus 00, 1\ominus 01,
1 \ominus 10, 00\ominus 11 \}$ of \emph{negative generating slopes}.
With this name at hand, we can see that the bottom reactions have the
common form
\begin{align}
  v & \rea v \oplus u \ominus v &
  u \ominus v \oplus u & \rea u
\end{align}
whenever $u$, $v \in \Sigma^*$ and $u \ominus v \in G_-$. This then
completes the compression of Table~\ref{tab:generator54}: The result
is Table~\ref{tab:generators54short}.

\begin{table}[t]
  \caption{The local reaction system for Rule 54, short form.}
  \centering
  \let\m=\multicolumn
  \tabcolsep=0pt
  \def\r{$&${}\rea}
  \def\e{$&${}=}
  \def\s{\phantom \sigma}
  \def\space{\hspace{1.8em}}
  \begin{tabular}{@{\space}rl@{\space\space}rl@{\space}}
    \toprule
    \m4c{Generating Slopes} \\
    \m4c{$G_- = \{\ominus 00, 1 \ominus 01, 1 \ominus 10, 00 \ominus 11\}$} \\
    \m4c{$G_+ = \{00\oplus, 01 \oplus 1, 01 \oplus 1, 11 \oplus 00\}$} \\
    \midrule
    \m4c{Reactions} \\
    $\ominus 00 \sigma \r \sigma \ominus 0\sigma$                & $\sigma 00 \oplus \r \sigma 0 \oplus$                        \\
    $\ominus 10 \s     \r \ominus 0$                             & $\s 01 \oplus     \r \s 0 \oplus$                            \\
    $\ominus 01 \sigma \r \bar\sigma \bar\sigma \ominus 1\sigma$ & $\sigma 10 \oplus \r \sigma 1 \oplus \bar\sigma \bar\sigma $ \\
    $\ominus 11 \sigma \r \bar\sigma \ominus 1\sigma$            & $\sigma 11 \oplus \r \sigma 1 \oplus \bar\sigma $            \\[1ex]
    $u \ominus v \oplus u \r u$ \\
    $v \r v \oplus u \ominus v$ & \m2l{for $u \ominus v \in G_-$} \\
    \midrule
    \m4c{Abbreviations} \\
    $\ominus 0 \e \ominus_1 0$ & $0 \oplus \e 0 \oplus_1$ \\
    $\ominus 1 \e \ominus_2 1$ & $1 \oplus \e 1 \oplus_2$ \\
    \bottomrule
  \end{tabular}
  \label{tab:generators54short}
\end{table}

\paragraph{Relation to the Transition Rule} In order to understand
this new form of the reaction system and to see how it is related to
the transition rule $\phi$, we write the reactions of $\Phi_-$ in the
following manner:
\begin{center}
  \def\c{\multicolumn1c}
  \def\s{\phantom{\sigma}}
  \def\undef{{\uparrow}}
  \begin{tabular}{l@{\qquad}lll}
    \toprule
    & \c{$\tau_0$} & \c{$\tau_1$} & \c{$\tau_2$} \\
    \midrule
    $\ominus_1 00 \sigma \rea \tau_0 \ominus_1 0\sigma$
    & $\phi(0, 0, \sigma) = \sigma$
    & $\phi(0, \sigma, \cdot)\undef$ \\
    $\ominus_2 10 \s \rea \ominus_1 0$
    & $\phi(0, \sigma, \cdot)\undef$ \\
    $\ominus_1 01 \sigma \rea \tau_0\tau_1 \ominus_2 1\sigma$
    & $\phi(0, 1, \sigma) = \bar\sigma$
    & $\phi(1, \sigma, \cdot) = \bar\sigma$
    & $\phi(\sigma, \cdot, \cdot)\undef$ \\
    $\ominus_2 11 \sigma \rea \tau_0 \ominus_2 1\sigma$
    & $\phi(1, \sigma, \cdot) = \bar\sigma$
    & $\phi(\sigma, \cdot, \cdot)\undef$ \\
    \bottomrule
  \end{tabular}
\end{center}
In the reactions at the leftmost column of the table, each variable
$\tau_i$ stands for the state of the cell at position $(0, i)$. The
other columns then show for each $\tau_i$ the computation that
determines its value -- or, if it cannot be computed, which
application of $\phi$ fails to have a determined value.

We can see e.\,g.\ in the first row that the state of the cell at $(0,
0)$ can be computed from the information presented in the initial
situation $\ominus_1 00 \sigma$. The cellular process of this
situation consists of the events $[-1, -1]0$, $[-1, 0]0$, and $[-1,
1]\sigma$, and therefore the state $\tau_0$ of the cell at $(0, 0)$
must be $\phi(0, 0, \sigma)$.

In the same way we can see that in the third row, $\tau_0$ is $\phi(0,
1, \sigma)$. The diagram contains however also entries for which not
all arguments of $\phi$ are known. The missing arguments are marked by
a dot. When the value of $\phi$ is independent of the missing
argument, it is entered in the table, otherwise the entry is marked
with an arrow.

We can see that the values of the $\tau_i$ only depend on three
equations,
\begin{equation}
  \label{eq:tau-equations}
  \phi(0, 0, \sigma) = \sigma, \qquad
  \phi(0, 1, \sigma) = \bar\sigma, \qquad
  \phi(1, \sigma, \cdot) = \bar\sigma\,.
\end{equation}
They all can be derived from the rule that a pair of touching 1's
cause a $\phi$ value of 0, while one or more isolated 1's make the
value equal to 1. In the case of $\phi(0, 0, \sigma)$, a pair of
touching 1's cannot occur, therefore the value of $\phi$ is one if and
only if $\sigma = 1$. In the other two cases, $\sigma = 1$ creates a
touching pair and $\sigma = 0$ inhibits it, therefore the function
value is $\bar\sigma$. In a similar way we can see that in the
remaining entries of the table, the value of $\phi$ is undefined. This
is how $\phi$ influences the reactions in $\Phi$.

In the table, the $\ominus$ have been written once again with
indices---not just to ease the translation from situations to cellular
processes, but also because with them we can see how many new events
are generated in the reactions. One can thus see that in the first
reaction one new event is generated because $\delta(\ominus_1 00
\sigma)$ must be equal to $\delta(\tau_0 \ominus_1 0 \sigma)$, and so
on. If the left side of a reaction has a $\ominus_i$ operator and the
right side a $\ominus_j$, then $j - i$ new cell states must be
generated in the reaction.

\paragraph{Slogans} These considerations may help to understand the
reactions of the system $\Phi$ a bit better. To help memorising them,
I will introduce two slogans. Both refer to the left side of the
reactions of $\Phi_-$. This side can always be written as $\ominus
\alpha\beta\sigma$, with $\alpha$, $\beta$, $\sigma \in\Sigma$. The
first slogan tells in which cases the value of $\alpha\beta$ makes the
reaction product longer or shorter than the initial term:
\begin{quote}
  ``01 causes growth, 10 shrinking, everything else no change.''
\end{quote}
The second slogan describes the influence of $\beta\sigma$ on the
newly generated cell states. They can be either be a copy ($\sigma$)
or the inversion ($\bar\sigma$) of the variable $\sigma$, and the rule
is:
\begin{quote}
  ``$0\sigma$ copies and $1\sigma$ inverts.''
\end{quote}

\section{Triangles and Ether}
\label{sec:triangles-ether}

In the rest of this article we will describe the behaviour of larger
systems of cells under Rule 54. We want to describe the interaction of
particles that move on a periodic background, the so-called
\emph{ether}. So we will now introduce, as a first step, reactions for
the ether. Since it has been done already to some extent in
\cite[Ch.~8]{Redeker2013a}, we will do it here in a shorter form and
from a higher point of view.

The first tool that we will use are \emph{reaction families}, which
allow to represent many similar reactions in a single formula.
Reaction families appeared already in\cite{Redeker2013a}, but here we
use a more streamlined notation.
\begin{definition}[Reaction Families]
  \label{def:reaction-family}
  If there is a reaction $a_k \rea b_k$ for every $k \geq 0$, we will
  write this as
  \begin{equation}
    \label{eq:reaction-family}
    (a_k \rea b_k)_k\,.
  \end{equation}
  The notation will be extended in the usual way to expressions like
  $(a_k \rea b_k)_{k \geq N}$ or $(a_{j,k} \rea b_{j,k})_{j,k}$. We
  will also speak of $(a_k)_k$ as a \emph{situation family}.
\end{definition}

\subsection{Triangle Reactions}

We will first find general formulas for reactions that represent
triangular structures like that in Figure~\ref{fig:example-reaction}.

There are two general laws that we will use here. The first one makes
it possible to \emph{iterate} a reaction of a special form. This can
be done in two ways,
\begin{subequations}
  \label{eq:iteration}
  \begin{align}
    \label{eq:right-iteration}
    \text{if}\quad
    a x \rea y a,
    &\qtext{then}
    (a x^k \rea y^k b)_k, \\
    \label{eq:left-iteration}
    \text{if}\quad
    x a \rea a y,
    &\qtext{then}
    (x^k a \rea b y^k)_k\,.
  \end{align}
\end{subequations}
The second law ``iterates'' a specific reaction family; in it, $n$ is
a constant:
\begin{equation}
  \label{eq:layer-iteration}
  \text{if}\quad
  (a_{k+n} \rea x a_k y)_k,
  \qtext{then}
  (a_{kn+i} \rea x^k a_i y^k)_{i, k}\,.
\end{equation}
Both laws can easily be proved by induction
\cite[Ch.~8.1]{Redeker2013a}.

We now search for cases in which the first law can be applied and in
which the left side is a generator reaction. There are two candidates,
$\ominus 000 \rea 0 \ominus 00$ and $\ominus 111 \rea 0 \ominus 11$.
The first one has $a = \ominus 00$ and $x = y = 0$ and leads to
\begin{subequations}
  \begin{equation}
    \label{eq:iterate-0}
    (\ominus 0^{k+2} \rea 0^k \ominus 00)_k,
  \end{equation}
  while the second reaction has $a = \ominus 11$, $x = 1$ and $y = 0$
  and leads to
  \begin{equation}
    \label{eq:iterate-1}
    (\ominus 1^{k+2} \rea 0^k \ominus 11)_k\,.
  \end{equation}
\end{subequations}

Family~\eqref{eq:iterate-0} is the more interesting one. It becomes
the core of another reaction family,
\begin{equation}
  \label{eq:101-layer}
  (1 0^{k+2} 1 \rea 10 \oplus 1 0^k 1)_k,
\end{equation}
whose derivation I will show here in detail, as an example for
calculation with reactions:
\begin{alignat*}{2}
  \underline{10}0 0^k 1
  &\rea 10 \oplus 1 \:\underline{\ominus\: 10} 00^k 1 \\
  &\rea 10 \oplus 1 \:\underline{\ominus\: 000^k} 1 \\
  &\rea 10 \oplus 1 0^k \:\underline{\ominus\: 001}
  \rea 10 \oplus 1 0^k 1 \ominus 01\,.
\end{alignat*}
Parts of the situations are underlined; they are the places that
change in the next reaction step. We will use this notation in later
calculation without special notice.

Reaction family~\eqref{eq:101-layer} can now be iterated by
rule~\eqref{eq:layer-iteration}, with $a_k = 1 0^k 1$ and $n = 2$. The
result is
\begin{equation}
  \label{eq:triangle-general}
  \left(1 0^{2k+i} 1
    \rea (10\oplus)^k 1 0^i 1 (\ominus 01)^k
  \right)_{i,k}\,.
\end{equation}
In families like these, the cases with $i < 2$ are the most important
ones, since the reactions in \eqref{eq:101-layer} have been applied in
them for the highest number of times. For $i = 1$, we can add one more
step, since we have $\underline{10}1 \rea 10 \oplus 1
\:\underline{\ominus\: 10} 1 \rea 10 \oplus 1 \ominus 01$.
therefore~\eqref{eq:triangle-general} can be written as two families,
\begin{subequations}
  \label{eq:triangle}
  \begin{align}
    \label{eq:triangle-even}
    \bigl(1 0^{2k} 1
    &\rea (10\oplus)^k 11 (\ominus 01)^k \bigr)_k,\\
    \label{eq:triangle-odd}
    \bigl(1 0^{2k+1} 1
    &\rea (10\oplus)^{k+1} 1 (\ominus 01)^{k+1} \bigr)_k\,.
  \end{align}
\end{subequations}
They, and all reactions of the form $(a_{k+n} \rea x^k a_n y^k)_k$,
are called \emph{triangle reactions}.

\begin{figure}[ht]
  \centering
  \def\arraystretch{1.5}
  \vspace{2ex}
  \begin{tabular}{c@{\hspace{3em}}c}
    \ci{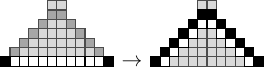} & \ci{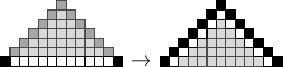} \\
    $\llap{$1 0^{6} 1$} \rea \rlap{$(10\op)^3 11 (\om 01)^3$}$ &
    $\llap{$1 0^{7} 1$} \rea \rlap{$(10\op)^4 1 (\om 01)^4$}$
  \end{tabular}
  \caption{Triangle reactions for $k=3$.}
  \label{fig:example-triangles}
\end{figure}

Diagrams for the reactions with $k = 3$ are shown in
Figure~\ref{fig:example-triangles}.

If we try the same manoeuvre with the other reaction family,
\eqref{eq:iterate-1}, we get $(0 1^{k+2} 0 \rea 01 \oplus 1 0^{k+2} 1
\ominus 10)_k$. This is a family to which~\eqref{eq:layer-iteration}
cannot be applied. Therefore we will now use the reaction
families~\eqref{eq:triangle} as our base for the description of the
ether.

\subsection{The Ether}

We will find now represent the ether of Rule 54 by reactions. The
reactions for Rule 54 will turn out to be a special case of a generic
scheme that applies to periodic patterns in any one-dimensional
cellular automaton.

In Rule 54 \cite{Boccara1991}, the ether is a periodic structure whose
configurations consist alternatingly of the two patterns $\dots
100010001 \dots$ and $\dots 011101110 \dots$\,. When one of them
occurs again, it is shifted horizontally by two cells, so that the
true time period is 4.

Our starting point for representing them by reactions must be the
configuration $\dots 100010001 \dots$, since to it we can apply one
reaction of type~\eqref{eq:triangle-odd},
\begin{equation}
  \label{eq:ether-triangle}
  10001 \rea (10 \oplus)^2 1 (\ominus 01)^2.
\end{equation}
It would be therefore advantageous to decompose the initial
configuration into components of the form 10001. With a small
extension of our notation, this is actually possible.

\begin{definition}[Overlapping Situations]
  \label{def:overlap}
  Let $a x$ be a situation. The $a \ovl{x}$ is also a situation, and
  $\ovl{x}$ is the \emph{overlapping part}. We have
  \begin{equation}
    \label{eq:overlap}
    \pr(a \ovl{x}) = \pr(a x)
    \qtext{and}
    \delta(a \ovl{x}) = \delta(a)\,.
  \end{equation}
  A product of situations with overlap, like $a \ovl{x} b \ovl{y}$, is
  only allowed if the situation $b y$ begins with $x$; then $a \ovl{x}
  b \ovl{y} = a b \ovl{y}$.

  A reaction that begins with $a \ovl{x}$ must have the form
  \begin{equation}
    \label{eq:overlap-reaction}
    a \ovl{x} \rea a' \ovl{x};
  \end{equation}
  it exists if $a x \rea a' x$ is a reaction.
\end{definition}

If we remind ourselves that the transitions of a cellular automaton
are defined in terms of overlapping cell neighbourhoods, then the new
extension looks quite natural.

We can now write a term like $(1000)^k 1$ as a product $(1000
\ovl{1})^k 1$ and apply the ether reactions in parallel to each
factor, except for the final 1. In this style,
Reaction~\eqref{eq:ether-triangle} is best written in the form $1000
\ovl1 \rea (10 \oplus \ovl1)^2 (1 \ominus 0 \ovl1)^2$.

But now we should better introduce abbreviations. We will write,
\begin{equation}
  \label{eq:epsilon-def}
  \epsilon_+ = 10 \oplus \ovl1
  \qqtext{and}
  \epsilon_- = 1 \ominus 0 \ovl1,
\end{equation}
such that~\eqref{eq:ether-triangle} becomes
\begin{equation}
  \label{eq:base->ether}
  1000 \ovl1 \rea \epsilon_+^2 \epsilon_-^2.
\end{equation}
The terms $\epsilon_+$ and $\epsilon_-$ are the simplest of the higher
level structures in Rule 54 that we will identify.

There is also a complementary reaction to~\eqref{eq:base->ether},
\begin{equation}
  \label{eq:ether->base}
  \epsilon_-^2 \epsilon_+^2 \rea 1000 \ovl1\,.
\end{equation}
In contrast to~\eqref{eq:base->ether}, this reaction does not belong
to a known family, and we will derive it by hand (see below). Together
the two reactions form a type that naturally represents the periodic
patterns of one-dimensional cellular automata. Before a formal
definition is given, we introduce the abbreviations
\begin{equation}
  \label{eq:background54}
  e_- = \epsilon_-^2, \qquad
  e_+ =\epsilon_+^2, \qquad
  b = 1000 \ovl1\,.
\end{equation}
Then we see that~\eqref{eq:base->ether} and~\eqref{eq:ether->base} are
example of the following general pattern:

\begin{definition}[Background Pairs]
  \label{def:background}
  Two situations, $e_-$, $e_+$, form a \emph{background pair} if there
  is a reaction
  \begin{subequations}
    \label{eq:all-background}
    \begin{equation}
      \label{eq:background}
      e_- e_+ \rea e_+ e_-\,.
    \end{equation}
    If there is also a situation $b \in \Sigma^*$ with
    \begin{equation}
      \label{eq:background-full}
      e_- e_+ \rea b \rea e_+ e_-,
    \end{equation}
  \end{subequations}
  then $b$ is the \emph{baseline} of the background pair.
\end{definition}

\begin{figure}[ht]
  \centering
  \begin{tikzpicture} [
    pin distance=2.5ex,
    every pin edge/.style={color=gray2, thin},
    reaction/.style={color=black, fill opacity=.2, draw=black,
      text opacity=1, text=black, auto, inner sep=0ex}]
    \def\xdist{0.6em}
    \tikzset{x=\xdist, y=\xdist}
    \def\tmax{18}
    \def\xmax{38}
    \coordinate (bottom left) at (-0.2, 0) {};
    \begin{scope}
      \clip (bottom left) rectangle (\xmax+.2, \tmax);
      \draw [gray3, very thin]
        let \n{both}={\xmax+\tmax} in
        foreach \x in {0, 2,..., \n{both}} {
          (\x, 0)         -- ++(-\tmax, \tmax)
          (\x - \tmax, 0) -- ++(\tmax, \tmax)
        };
    \end{scope}
    \draw[gray2, thick] (bottom left) node[black, left, font=\scriptsize] {$t = 0$}
        -- (\xmax+.2, 0);
    \draw[reaction] (4, 14)
        -- coordinate [midway, pin=-170:$e_-$] ++(1, -1);
    \draw[reaction] (4, 6)
        -- coordinate [midway, pin=170:$e_+$] ++(1, 1);
    \draw[reaction, thick] (4, 0)
        -- coordinate [midway, pin=140:$b$] ++(2, 0);
    \fill[reaction] let \n1={6} in
       (10, 0)
        -- node[midway] {$e_+^{\n1}$} ++(\n1, \n1)
        -- node[midway] {$e_-^{\n1}$} ++(\n1, -\n1)
        -- node[midway, below=.4] {$b^{\n1}$} cycle;
    \fill[reaction] let \n1=7, \n2=4 in
       (24, 8)
        -- node[midway] {$e_+^{\n1}$} ++(\n1, \n1)
        -- node[midway] {$e_-^{\n2}$} ++(\n2, -\n2)
        -- node[midway] {$e_+^{\n1}$} ++(-\n1, -\n1)
        -- node[midway] {$e_-^{\n2}$} cycle;
  \end{tikzpicture}
  \caption{An ether, represented by a background pair $e_-$, $e_+$
    with baseline~$b$.}
  \label{fig:ether-tiling}
\end{figure}
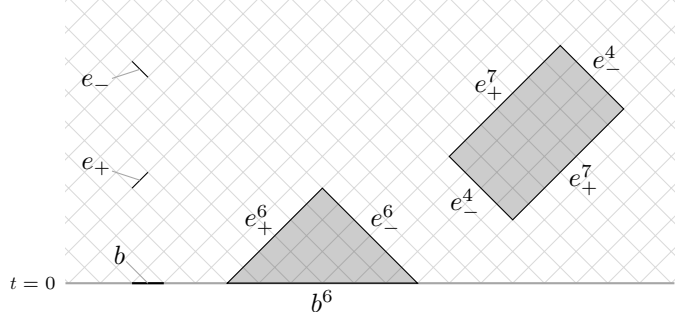

A background pairs represent the elementary region of a tiling of the
two-dimensional space-time (Figure~\ref{fig:ether-tiling}).
If a background pair is present, we automatically get the reaction
families
\begin{subequations}
  \begin{align}
    \label{eq:baseline-repeated}
    (b^k &\rea e_+^k e_-^k)_k,\\
    \label{eq:ether-repeated}
    (e_-^k e_+^\ell &\rea e_+^\ell e_-^k)_{k,\ell},
  \end{align}
\end{subequations}
which represent larger patches of the background. As we can see in
Figure~\ref{fig:ether-tiling}, the reactions
of~\eqref{eq:baseline-repeated} represent the generation of a larger
piece of ether from an initial configuration,
while~\eqref{eq:ether-repeated} represents the development of a
background fragment at a later time.

\paragraph{Derivation of the remaining ether reaction}
\label{sec:derivations} We have not yet proved
equation~\eqref{eq:ether->base}, the reaction $e_- e_+ \rea 1000
\ovl1$. This will be done now.

The computation is an example for a larger calculation with Flexible
Time. We will prove~\eqref{eq:ether->base} via the two reactions
\begin{subequations}
  \label{eq:resolution-steps}
  \begin{align}
    \label{eq:epsilon-resolve}
    \epsilon_- \epsilon_+ &\rea 1^2 \ovl1, \\
    \label{eq:epsilon-pair-resolve}
    \epsilon_-^2 \epsilon_+^2 &\rea 1000 \ovl1 \\
  \intertext{and the auxiliary step}
    \label{eq:epsilon-auxiliary}
    0 1^3 0 &\rea 01 \oplus 1 0^3 1 \ominus 10\,.
  \end{align}
\end{subequations}

The last reaction is an element of the reaction family
\begin{equation}
  \label{eq:010-layer}
  (0 1^{k+2} 0 \rea 01 \oplus 1 0^{k+2} 1 \ominus 10)_k\,.
\end{equation}
Its derivation uses the reaction family~\eqref{eq:iterate-1} and is
done in the following way:
\begin{align*}
  \underline{01}1 1^k 0
  &\rea 01 \oplus 1 \:\underline{\ominus\: 011} 1^k 0 \\
  &\rea 01 \oplus 1 \:\underline{\ominus\: 11 1^k} 0 \\
  &\rea 01 \oplus 100 0^k \:\underline{\ominus\: 11 0}
  \rea 01 \oplus 1 0^{k+2} 1 \ominus 10\,.
\end{align*}
Now we can derive the other two reactions
of~\eqref{eq:resolution-steps}:
\begin{align*}
  \epsilon_- \epsilon_+
  &= 1 \ominus 0 \ovl1 1 0 \oplus \ovl1 \\
  &= 1 \:\underline{\ominus\: 0 1} 0 \oplus \ovl1 \\
  &\rea 11 \underline{1 \ominus 10 \oplus \smash{\ovl1}}
  \rea 11 \ovl1, \\
  \epsilon_- \underline{\epsilon_- \epsilon_+} \epsilon_+
  &\rea \epsilon_- 11 \ovl1 \epsilon_+ \\
  &= 1 \ominus 0 \ovl1 11 \ovl1 10 \oplus \ovl1 \\
  &= 1 \ominus \underline{0 1^3 10} \oplus \ovl1 \\
  &\rea \underline{1 \ominus 01 \oplus 1} 0^3
  \underline{1 \ominus 01 \oplus \smash{\ovl1}}
  \rea 1 0^3 \ovl1\,.
\end{align*}
In the second computation we have used~\eqref{eq:epsilon-resolve}
and~\eqref{eq:epsilon-auxiliary}.

\section{Particles}
\label{sec:particles}

\begin{figure}[ht]
  \centering
  \tabcolsep=2pt
  \def\arraystretch{1.3}
  \begin{tabular}{cccc}
    \ci{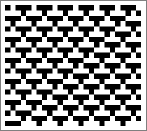} & \ci{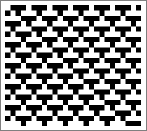} & \ci{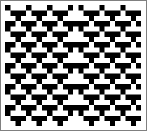} & \ci{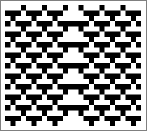} \\
    $\wr$ & $\wl$ & $g_o$ & $g_e$
  \end{tabular}
  \caption{Particles under Rule 54. The diagrams show the four types
    of gliders on an ether background.}
  \label{fig:gliders}
\end{figure}

In the ether particles move. Boccara \emph{et al.}\ \cite{Boccara1991}
have found four of them and called them $\wl$, $\wr$, $g_o$ and $g_e$
(Figure \ref{fig:gliders}). We will refer to the moving particles
$\wl$ and $\wr$ sometimes as \emph{gliders}, in contrast to the static
particles $g_o$ and $g_e$.

Now we will represent these particles by situations and reactions. The
characterisation of particles is a natural generalisation of that of a
background:
\begin{definition}[Particles]
  \label{def:particles}
  Let $(b_-, b_+)$ be a background pair. A \emph{particle} that moves
  in this background is a situation $p$ for which there is a reaction
  \begin{equation}
    \label{eq:particle}
    b_-^m p b_+^n \rea b_+^n p b_-^m\,.
  \end{equation}
  The pair $(m, n)$ is the \emph{type} of the particle.
\end{definition}

\begin{figure}[ht]
  \centering
  \begin{tikzpicture}
    \def\xdist{2em}
    \tikzset{x=\xdist, y=\xdist}
    \def\mval{2}  \def\nval{3}
    \coordinate (particle) at (1.5, 0);
    \path (0, 0) coordinate (p0l)
       -- ++(particle) coordinate (p0r)
       -- ++(\nval, \nval)   coordinate (right)
       -- ++(-\mval, \mval)  coordinate (p1r)
       -- ++($-1*(particle)$) coordinate (p1l)
       -- ++(-\nval, -\nval)  coordinate (left)
       -- cycle;
    \path let \n0={4.25} in
       (p0l) -- ++(-\n0,0) coordinate (sw)
       (p1r) -- ++(\n0,0)  coordinate (ne)
       coordinate (nw) at (sw |- ne)
       coordinate (se) at (ne |- sw);
    \begin{scope}
      \clip (nw) -- (p1l) -- (left) -- (p0l) -- (sw) -- cycle;
      \draw[help lines, ]
        let \n1={\mval+\nval}, \n2={\n1+6} in {
           \foreach \i in {0, 2, ..., \n2} {
             { [shift=(p0l)] (2*\nval-\i, 0) -- ++(-\n1, \n1) }
             { [shift=(p1l)] (2*\mval-\i, 0) -- ++(-\n1, -\n1) }
           }
         };
    \end{scope}
    \begin{scope}
      \clip (ne) -- (p1r) -- (right) -- (p0r) -- (se) --cycle;
      \draw[help lines, shift=(p0r)]
        let \n1={\mval+\nval}, \n2={\n1+6} in
          \foreach \i in {0, 2, ..., \n2} {
            (\i-2*\mval, 0) -- ++(\n1, \n1)
            (\i, 0) -- ++(-\n1, \n1)
          };
    \end{scope}
    \draw[help lines, text=black, inner sep=0ex]
      (left) -- node[above right] {$b_-^\mval$} (p0l)
      (left) -- node[below right] {$b_-^\nval$} (p1l)
      (p0r) -- node[above left, pos=0.28] {$b_-^\nval$} (right)
      (p1r) -- node[below left] {$b_-^\mval$} (right);
    \draw[very thick]
      (p0l) -- node[below] {$p$} (p0r)
      (p1l) -- node[below] {$p$} (p1r);
    \draw[->, densely dashed]
      (p0l) -- node[right] {$(\Delta t, \Delta x)$} (p1l);
  \end{tikzpicture}
  \caption{A particle of type $(2, 3)$ as part of a periodic
    background. Its relative speed is $\frac{1}{5}$.}
  \label{fig:particle}
\end{figure}
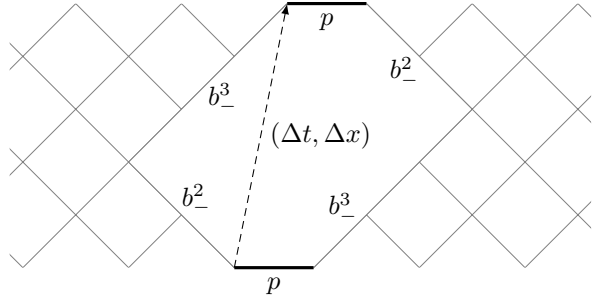

The type of $p$ represents its speed relative to the background. To
convert it to a more conventional form, we notice that in the initial
situation of the reaction~\eqref{eq:particle}, the left side of $p$ is
located at the space-time point $m \delta(b_-)$, while in its final
situation, it is at $n \delta(b_+)$. The \emph{period vector} $(\Delta
t, \Delta x) = n \delta(b_+) - m \delta(b_-)$ is therefore the
displacement that $p$ undergoes during one cycle of its existence.
After $\Delta t$ time steps, the particle is in the same state, and it
has $\Delta x$ positions to the right. The speed of $p$ is
then~$\frac{\Delta x}{\Delta t}$. (Figure \ref{fig:particle}.)

Often it is simpler to work with speeds relative to the background.
For this we use the vectors $T = \delta(b_+) - \delta(b_-)$ and $X =
\delta(b_+) + \delta(b_-)$ as our base, the first one pointing to the
future and the second one to the right. A particle of type $(m, n)$
has then a period vector of $\frac{n + m}{2} T + \frac{n - m}{2} X$
and we can say that its \emph{relative speed} is $\frac{n - m}{n +
  m}$.

\paragraph{The particles of Rule 54} For Rule 54 we use the following
definitions:
\begin{align}
  \label{eq:particles54}
  \wl &= \epsilon_- 1^2 \ovl{1}, & g_o &= \epsilon_+ \epsilon_-, \notag\\
  \wr &= 1^2 \epsilon_+,         & g_e &= \epsilon_+ 1 \epsilon_-\,.
\end{align}
They have this specific form because we can then use a simple subset
of our reaction system to represent their behaviour. This subset
consists of two reaction families and one extra reaction,
\begin{subequations}
  \label{eq:epsilon-triangle}
  \begin{align}
    \label{eq:epsilon-triangle-even}
    (\epsilon_- 1^{2k} \epsilon_+
    &\rea \epsilon_+^{k+1} \epsilon_-^{k+1})_{k \geq 1},
    & \epsilon_- \epsilon_+ &\rea 1^2 \ovl{1},
    \\
    \label{eq:epsilon-triangle-odd}
    (\epsilon_- 1^{2k+1} \epsilon_+
    &\rea \epsilon_+^{k+1} 1 \epsilon_-^{k+1})_k,
  \end{align}
\end{subequations}
which transform situations that consist only of $\epsilon_-$,
$\epsilon_+$ and 1 into each other. They can easily be derived from
the reaction families~\eqref{eq:010-layer} and~\eqref{eq:triangle}.
With the reactions of~\eqref{eq:epsilon-triangle-even}, the ether
reaction $\epsilon_- \epsilon_+$ can be proved, as we have seen on
page~\pageref{sec:derivations}.

\begin{figure}[ht]
  \centering
  \def\arraystretch{1.5}
  \begin{tabular}{c}
    \ci{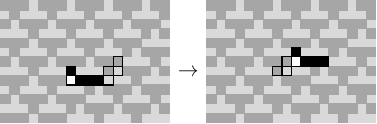}\\
    $\wr e_+ \rea e_+ \wr$\\[4ex]
    \ci{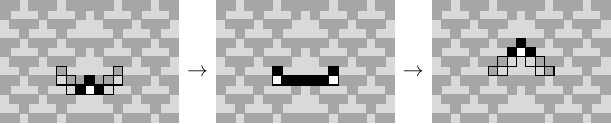}\\
    $e_- g_o e_+
    \rea \wr \wl
    \rea e_+ g_o e_-$.\\[4ex]
    \ci{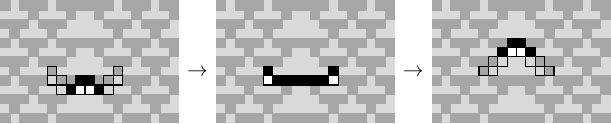}\\
    $e_- g_e e_+
    \rea \wr 1 \wl
    \rea e_+ g_e e_-$.
  \end{tabular}
  \caption{Evolution of the Rule 54 particles. The particles are shown
    in strong colours, and the outlined squares are ether.}
  \label{fig:particle-movements}
\end{figure}

With these reactions we can now verify that the terms
in~\eqref{eq:particles54} are indeed particles:
\begin{subequations}
  \label{eq:particle-reactions}
  \begin{alignat}{4}
    \label{eq:wr-reaction}
    \wr e_+
    &= \epsilon_- 1^2 \epsilon_+^2
    \rea \epsilon_+^2 \epsilon_-^2 \epsilon_+
    \rea \epsilon_+^2 \epsilon_- 1^2 \ovl{1}
    &&= e_+ \wr,\\
    \label{eq:go-reaction}
    e_- g_o e_+
    &= \epsilon_-^2 \epsilon_+ \epsilon_- \epsilon_+^2
    \rea \epsilon_- 1^2 1^2 \epsilon_+
    \rea \epsilon_+^3 \epsilon_-^3
    &&= e_+ g_o e_-,\\
    \label{eq:ge-reaction}
    e_- g_e e_+
    &= \epsilon_-^2 \epsilon_+ 1 \epsilon_- \epsilon_+^2
    \rea \epsilon_- 1^2 1 1^2 \epsilon_+
    \rea \epsilon_+^3 1 \epsilon_-^3
    &&= e_+ g_e e_- \,.
  \end{alignat}
\end{subequations}
The reaction $e_- \wl \rea \wl e_-$ has been omitted since the
reactions in~\eqref{eq:particles54} are left-right symmetric. We see
from these reactions that the types of $\wr$ and $\wl$ are $(0, 1)$
and $(1, 0)$, while $g_o$ and $g_e$ both have type $(1, 1)$.
Figure~\ref{fig:particle-movements} contains diagrams of the
reactions.

\paragraph{Collisions of two particles} With the reactions
of~\eqref{eq:epsilon-triangle} we can already find out simple facts
about the particles and their interactions. One fact is hidden
in~\eqref{eq:go-reaction}: the reaction
\begin{equation}
  \label{eq:w+w->go}
  \wr \wl \rea e_+ g_o e_-
\end{equation}
can easily be recognised once we remember that $\wr \wl = \epsilon_-
1^2 1^2 \epsilon_+$. This is the reaction in which two colliding $w$
particles create a $g_o$. It is in fact the only reaction that is
possible between the two $w$ particles. To see this, we note that if
$\wr$ moves towards $\wl$ with nothing else than ether between them,
this must be represented by a situation $\wr E \wl$, where $E$ is a
product of an arbitrary number of $e_-$ and $e_+$ terms. Then there
must be a reaction $E \rea e_+^m e_-^n$, where $m$ is the number of
$e_+$ factors in $E$ and $n$ the number of $e_-$ factors. This leads
to a reaction chain
\begin{equation}
  \label{eq:w-collision-chain}
  \wr E \wl
  \rea \wr e_+^m e_-^n \wl
  \rea e_+^m \wr \wl e_-^n
\end{equation}
to which we can apply~\eqref{eq:w+w->go}. We have thus seen that two
$w$ gliders always move towards each other unchanged until they react
to the position $\wr \wl$, and that therefore~\eqref{eq:w+w->go} is
their only possible collision.

The same principle can be applied to any pair of colliding particles.
We have then the following theorem:
\begin{theorem}[Particle Collisions]
  \label{thm:particle-collisions}
  Let $p$ and $p'$ be two particles of types $(m, n)$ and $(m', n')$,
  with $p$ left of $p'$. Then $p$ moves toward $p'$ if $n m' > m n'$,
  away from $p'$ if $n m' < m n'$, otherwise they keep the same
  distance.

  If they collide, then there are $n m'$ possible interactions between
  them.
\end{theorem}

\begin{proof}
  If $p$ and $p'$ collide, the relative speed of $p$ must be greater
  than that of~$p'$. This means that $\frac{n - m}{n + m} > \frac{n' -
    m'}{n' + m'}$, or equivalently that $n m' > m n'$. The other two
  cases are similar.

  For the second statement we represent the relative positions of $p$
  and $p'$ by a situation $a p b p' c$ with $a$, $b$, $c \in \{ b_-,
  b_+ \}^*$. Here $a$ and $c$ represent the empty space left and right
  of the particles. We can make them arbitrarily large without changing
  the relative position of $p$ and $p'$. (A change of $a$ changes the
  \emph{absolute} position of $p$ and $p'$, but that has no influence
  on their behaviour.) Especially we can assume that $a = b_-^m$ and
  $c = b_+^{n'}$. The situation $b$ represents the space between $p$
  and $p'$, and we can always bring it by background reactions to the
  form $b_+^i b_-^j$.

  So we can assume that the environment of the particles has the form
  $b_-^m p b_+^i b_-^j p' b_+^{n'}$. Since $p$ and $p'$ collide, none
  of the reactions $b_-^m p b_+^n \rea b_+^n p b_-^m$ and $b_-^{m'} p'
  b_+^{n'} \rea b_+^{n'} p' b_-^{m'}$ can be applied to this
  situation. This means that $i < n$ and $j < m'$, for which there are
  $n m'$ possibilities.
\end{proof}

\paragraph{Interaction between the static particles and the
  \boldmath$w$ gliders.} When we start with a random initial
configuration and let it evolve for a short time, we typically see
some $g_o$ and $g_e$ particles on a background, with $\wr$ and $\wl$
moving between them (Figure~\ref{fig:random}). The formalism for Rule
54 is now developed far enough to describe with it the behaviour of
these particles in reasonable detail.

Specifically, we can now describe the behaviour of \emph{isolated}
$g_o$ and $g_e$ particles, which never interact with each other, only
with $\wr$ and $\wl$. In Flexible Time we can express this requirement
by restricting ourselves to the reactions that start from a situation
$x g y$ with $x \in \{ e_-, \wr \}^*$, $g \in \{ g_o, g_e \}$ and $y
\in \{ e_+, \wl \}^*$.

The $g_o$ case is the simplest, since the collision with a $w$ always
destroys this particle. Up to symmetry we have only the following
reactions,
\begin{equation}
  \label{eq:w+g_o}
  \wr g_o e_+ \rea e_+ \wr e_-,
  \qquad
  \wr g_o \wl \rea e_+^2 e_-^2\,.
\end{equation}
They could be verified directly, but we will now compute them in a way
that is also useful in the more complex case of $g_e$. For this we
begin with $\wr g_o$, a common factor of the two left sides
in~\eqref{eq:w+g_o}, and also the smallest situation that represents a
collision of $\wr$ and $g_o$. Their reaction is $w g_o =
\underline{\epsilon_- 1^2 \epsilon_+} \epsilon_- \rea \epsilon_+^2
\epsilon_-^3 = e_+ \epsilon_- e_-$. The end result is here interpreted
as an $\epsilon_-$ surrounded by two ether fragments. We can consider
it as a short-lived intermediate stage, or a \emph{resonance}, if we
use once again the jargon of particle physics. In the next step we
ignore the ether fragments and consider only the development of the
$\epsilon_-$. There are two ways in which it can interact with an
ether fragment or a $w$ particle, namely through the reactions
$\epsilon_- e_+ = \epsilon_- \epsilon_+^2 \rea 1^2 \epsilon_+ = \wl$
and $\epsilon_- \wr = \epsilon_- 1^2 \epsilon_+ \rea \epsilon_+^2
\epsilon_-^2 = e_+ e_-$. No further resonances arise from these
reactions, so we can stop here.

The result is a scheme of three reactions; they describe the behaviour
of $g_o$ in the same way as~\eqref{eq:w+g_o}:
\begin{subequations}
  \label{eq:w+go-full}
  \begin{align}
    \label{eq:go-collision}
    \wr g_o        & \rea e_+ \epsilon_- e_-, \\[1ex]
    \epsilon_- e_+ & \rea \wr \notag          \\
    \epsilon_- \wl & \rea e_+ e_-
    \label{w+go-decay}
  \end{align}
\end{subequations}
One can use them to derive the reactions of~\eqref{eq:w+go-full},
e.\,g.\ with the reaction chain $\underline{\wr g_o} e_+ \rea e_+
\epsilon_- \underline{e_- e_+} \rea e_+ \underline{\epsilon_- e_+} e_-
\rea e_+ \wl e_-$ for the first reaction. But for most purposes,
\eqref{eq:w+go-full} can be interpreted directly as a two-step scheme
that describes how an $\epsilon_-$ is created~\eqref{eq:go-collision}
and how it decays to $\wr$ or ether~\eqref{w+go-decay}. The ether
particles at the right side of~\eqref{eq:go-collision} can be thought
as becoming part of the surrounding space, which is why they do not
appear in~\eqref{w+go-decay}.

A similar but more complex scheme describes the collision of $g_e$
with one or more $w$ particles. Up to symmetry it has the intermediate
states 1, $1 \epsilon_-$ and $1^5 \ovl1$ and can be written as
follows:
\begin{subequations}
  \label{eq:ge-full}
  \begin{align}
    \label{eq:ge-collision}
    \wr g_e          & \rea e_+ e_- 1 \epsilon_-   \\[1ex]
    1 \epsilon_- e_+ & \rea 1 \wl \notag             \\
    1 \epsilon_- \wl & \rea 1 e_+ e_-
    \label{eq:ge-1-epsilon} \\[1ex]
    e_- 1 e_+        & \rea 1^5 \ovl1 \notag         \\
    e_- 1 \wl        & \rea \wl 1 e_- \notag         \\
    \wr 1 \wl        & \rea e_+ g_e e_-
    \label{eq:ge-1} \\[1ex]
    e_- 1^5 e_+      & \rea \wl g_e \wr \notag       \\
    e_- 1^5 \wl      & \rea \wl e_+^2 1 e_-^2 \notag \\
    \wr 1^5 \wl      & \rea e_+^2 g_e e_-^2.
    \label{eq:ge-1-5}
  \end{align}
\end{subequations}
All these reactions are short and can be verified directly. They show
that an isolated $g_e$ can neither be destroyed nor does it explode to
a larger structure. (See \cite{Martin2000} for the deeper reasons
behind this.) The intermediate states can however persist for an
indefinite time if the right pattern of incoming $w$ gliders is given.
One can see this e.\,g.\ from the reaction $e_- 1 \wl \rea \wl 1 e_-$
in~\eqref{eq:ge-1}. It can be iterated to $(e_-^k 1 \wl^k \rea \wl^k 1
e_-^k)_k$, which shows how the intermediate state 1 can be kept alive
indefinitely by a sequence of incoming $\wl$ gliders.

In summary we get a description of the behaviour not just of a single
$g_o$ and $g_e$, but also of a whole system of particles, provided
that the $g$ particles and their intermediate states all keep a
distance from each other. The distance must be so large that next to
each $g$ particle or intermediate state there is always a $w$ particle
or an ether fragment. As long as this is true, the $g_o$ particles are
created~\eqref{eq:w+w->go} and destroyed~\eqref{eq:w+go-full} by $w$
gliders, while the $g_e$ persist but go through intermediate
states~\eqref{eq:ge-full}.

\section{Summary}
\label{sec:conclusion}

This text consists of two interleaving tracks, one with the goal of
understanding Rule 54 better, the other to find concepts that are
valid for all cellular automata.

After a recapitulation of the results derived in \cite{Redeker2013a},
we began with constructing a shorter representation of the local
reaction system for Rule 54 (Table~\ref{tab:generators54short}). We
then described how the transition rule $\phi$ influences the local
reaction system $\Phi$ and at the end introduced two slogans to
summarise the generator reactions of the local system.

With~\eqref{eq:iteration} and~\eqref{eq:layer-iteration}, we learned
how to iterate reactions. This helped to derive expressions for the
triangles under Rule 54 and to find a
subsystem~\eqref{eq:epsilon-triangle} of $\Phi$ that consists only of
modified triangle reactions. It also introduced the situations
$\epsilon_-$ and $\epsilon_+$, which, together with the situation 1,
were the building blocks of the following construction.

We introduced definitions for the background and for particles and
explored particle collisions. A formula for the number of particle
interactions was already found in \cite{Hordijk2001} under a different
framework, but the proof here seems more direct.

Expressions for the ether and the main particles of Rule 54 were found
and the collisions of the particles computed. We could see that an
isolated $g_e$ is stable under all collisions with incoming $w$
gliders. This extends in a way a result in \cite{Martin2000}, which
already showed that a single $g_e$ could not be destroyed, but the
current, more detailed investigation also shows that it could not
``explode'' either and become a steadily growing perturbation in the
ether.

On the way to this result, we saw an efficient method to display all
possible interactions of an isolated particle with all other particles
and the background~\eqref{eq:ge-full}.

The track about Rule 54 lead therefore to results about the
interaction of its particles, while the general track lead to generic
definitions of triangles, background and particles and a theorem about
glider collisions. Both show how Flexible Time helps to understand an
automaton like Rule 54 as a system of interacting particles.

\paragraph{Changes in the formalism} One of the aims of this work was
to extend the capabilities of Flexible Time by applying it to the
understanding of a ``naturally occuring'' cellular automaton, i.\,e.\
one that was not constructed for a specific purpose. This resulted in
the following changes with respect to the version in
\cite{Redeker2013a}:
\begin{enumerate}
\item The interpretation of $\ominus$ and $\oplus$ were changed
  silently in~\eqref{eq:indices}. In \cite{Redeker2013a}, they were
  abbreviations for $\ominus_r$ and $\oplus_r$, where $r$ was the
  radius of the cellular automaton. Now the horizontal offsets
  associated to $\ominus$ and $\oplus$ depend on the context in which
  the symbols occur.

\item Reaction families, which were already present in
  \cite{Redeker2013a}, got a shorter notation.
\item A short notation for overlapping situations was introduced in
  Definition~\ref{def:overlap}. There was already an overlap notation
  in \cite{Redeker2013a}, but it was more clumsy. Now overlapping
  situations are part of the normal formalism.
\end{enumerate}

The new interpretation of $\ominus$ and $\oplus$ allowed us to write
the formulas of the local reaction system completely without indices
and to make the similarities between the basic reactions more visible.

With overlaps, definitions like those of a background
pair~\eqref{eq:all-background} could be written in a concise way.

\paragraph{Acknowledgement} I want to thank Nazim Fatès for reading the
manuscript and giving many helpful hints.

\bibliography{../references}

\end{document}